\documentclass[aps,twocolumn,superscriptaddress,amsmath,amssymb]{revtex4-2}
\usepackage{amssymb}
\usepackage{adjustbox}
\usepackage{lipsum}
\usepackage{multirow}
\usepackage{amsmath}
\usepackage{graphics}
\usepackage{color}
\usepackage{natbib}

\newcommand{\be}{\begin{equation}}
\newcommand{\ee}{\end{equation}}
\newcommand{\bea}{\begin{eqnarray}}
\newcommand{\eea}{\end{eqnarray}}
\newcommand{\bse}{\begin{subequations}}
\newcommand{\ese}{\end{subequations}}
\newcommand{\smb}{${\rm SmMg_2Bi_2}$}

\begin{document}
\title{Electronic and magnetic properties of the topological semimetal SmMg$_2$Bi$_2$}
\author{Asish K. Kundu}
\email{akundu@bnl.gov}
\affiliation{Condensed Matter Physics and Materials Science Division, Brookhaven National Laboratory, Upton, New York 11973, USA}
\author{Santanu Pakhira}
\affiliation{Ames National Laboratory, Iowa State University, Ames, Iowa 50011, USA}
\author{Tufan Roy}
\affiliation{Research Institute of Electrical Communication, Tohoku University, Sendai 980-8577, Japan}
\author{T. Yilmaz}
\affiliation{National Synchrotron Light Source II, Brookhaven National Laboratory, Upton, New York 11973, USA}
\author{Masahito Tsujikawa}
\affiliation{Research Institute of Electrical Communication, Tohoku University, Sendai 980-8577, Japan}
\author{Masafumi Shirai}
\affiliation{Research Institute of Electrical Communication, Tohoku University, Sendai 980-8577, Japan}
\affiliation{Center for Science and Innovation in Spintronics, Core Research Cluster, Tohoku University, Sendai 980-8577, Japan}
\author{E. Vescovo}
\affiliation{National Synchrotron Light Source II, Brookhaven National Laboratory, Upton, New York 11973, USA}
\author{D. C. Johnston}
\affiliation{Ames National Laboratory, Iowa State University, Ames, Iowa 50011, USA}
\affiliation{Department of Physics and Astronomy, Iowa State University, Ames, Iowa 50011, USA}
\author{Abhay N. Pasupathy}
\affiliation{Condensed Matter Physics and Materials Science Division, Brookhaven National Laboratory, Upton, New York 11973, USA}
\affiliation{Department of Physics, Columbia University, New York, 10027, USA}
\author{Tonica Valla}
\affiliation{Condensed Matter Physics and Materials Science Division, Brookhaven National Laboratory, Upton, New York 11973, USA}
\affiliation{Donostia International Physics Center (DIPC), P. Manuel de Lardizabal 4, 20018 San Sebastián, Spain}
\date{\today}

\begin{abstract}

Dirac semimetals show nontrivial physical properties and can host exotic quantum states like Weyl semimetals and topological insulators under suitable external conditions. Here, by combining angle-resolved photoemission spectroscopy measurements (ARPES) and first-principle calculations, we demonstrate that the Zintl-phase compound SmMg$_2$Bi$_2$ is in close proximity to a topological Dirac semimetallic state. ARPES results show a Dirac-like band crossing at the zone-center near the Fermi level ($E_\mathrm {F}$) which is further confirmed by first-principle calculations. Theoretical studies also reveal that SmMg$_2$Bi$_2$ is a $Z_2$ topological class and hosts spin-polarized states around the $E_\mathrm {F}$. Zintl's theory predicts that the valence state of Sm in this material should be Sm$^{2+}$, however we detect many Sm-4$f$ multiplet states (flat-bands) whose energy positions and relative intensities suggest the presence of both dominant Sm$^{2+}$ and minor Sm$^{3+}$. The small concentration (2.5\%) of Sm$^{3+}$ in the bulk of a crystal is inferred to arise from Sm vacancies in the crystal. It is also evident that these flat-bands and other dispersive states are strongly hybridized when they cross each other. Due to the presence of Sm$^{3+}$ ions, the temperature dependence of the magnetic susceptibility $\chi(T)$ shows a Curie-Weiss-like contribution in the low temperature region, in addition to the Van Vleck-like behavior expected for the Sm$^{2+}$ ions. The present study will help to better understand the electronic structure, magnetism, and transport properties of related materials.

\end{abstract}


\maketitle

\section{Introduction}
Three-dimensional (3D) topological Dirac semimetals (TDSs) are analogous to 2D graphene, with valence and conduction bands touching only at discrete points close to Fermi energy ($E_{\rm F}$) in the Brillouin zone (BZ) and disperse linearly in all directions. A similar scenario can be observed at the quantum-critical point of a topological phase transition between a 3D topological insulator and a normal insulator at a gap-inversion point \cite{yang2014classification}. A 3D TDS itself is the parent state of 3D topological insulators, which become either a 3D strong topological insulator or a 3D topological crystalline insulator when additional symmetry is broken \cite{PhysRevB.85.195320,PhysRevB.88.125427}. The stable 3D Dirac semimetallic phase was initially observed in Na$_3$Bi \cite{liu2014discovery} and Cd$_3$As$_2$ \cite{liu2014stable,neupane2014observation} compounds, and now extends to other material classes \cite{takane2021dirac,anh2021elemental,ma2020emergence}. Recently, a TDS phase was predicted for many intermetallic compounds with Zintl phases \cite{feng2022prediction,ma2020emergence} and indeed detected experimentally in BaMg$_2$Bi$_2$ \cite{takane2021dirac}. Zintl-phase compounds have been recently reported to exhibit rich electronic, magnetic, and transport properties associated with nontrivial band topology, such as high carrier mobility, giant non-saturating magnetoresistance, topological and anomalous Hall effects, and robust surface states \cite{rosa2020colossal,c2008zintl,yan2022field,kondo2022field,kundu2022topological,xu2019higher,niu2019quantum,li2019dirac,PhysRevB.104.174427,kundu2021quantum,chang2019realization,pakhira2020magnetic,marshall2021magnetic,cao2022giant}. Additionally, many Zintl compounds display superior thermoelectric (TE) and magnetic properties, which make them suitable for a wide range of applications, including power generation, waste-heat conversion, and solid-state Peltier coolers \cite{disalvo1999thermoelectric,he2017advances,gascoin2005zintl,pakhira2021zero}.

\begin{figure*}[ht!]
\centering
\includegraphics[width=17cm]{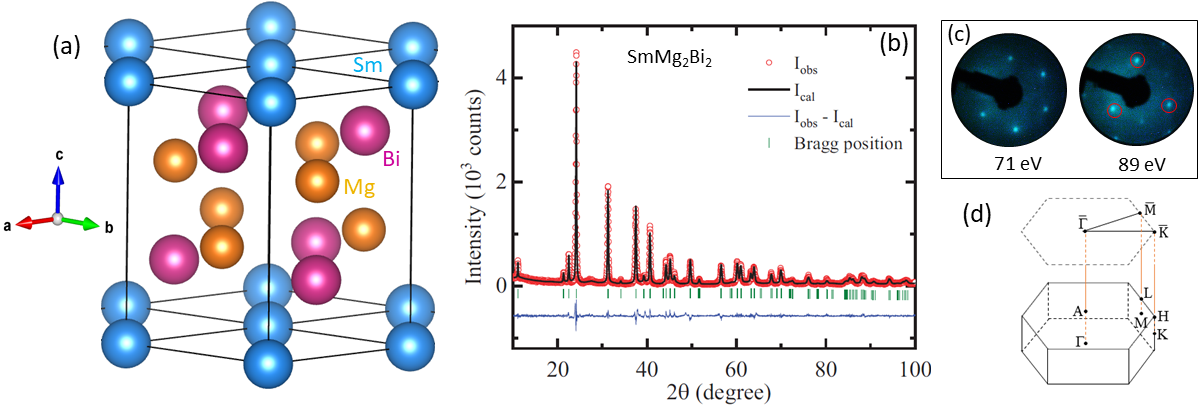}
\caption {(a) Crystal structure of trigonal SmMg$_2$Bi$_2$. (b) XRD patterns of powdered SmMg$_2$Bi$_2$, along with the refinement of the data using FULLPROF software. (c) Hexagonal LEED patterns of SmMg$_2$Bi$_2$ single crystals recorded at energies 71 eV and 89 eV. At 89 eV, only three spots are clearly visible (red circles) and other three are faint, reflecting the 3-fold rotational symmetry of the crystal. (d) Schematics of the hexagonal bulk Brillouin zone (bottom) and its surface projection (SBZ, top).}\label{Fig1}
\end{figure*}

Among Zintl families, $AB_2X_2$ ($A$ = alkaline earth metals and rare-earths; $B$ = Zn, Mn, Cd, Mg; $X$ = Sb, Bi, As) type compounds have gained more attention due to their tunable electronic and thermoelectric properties through the fine-tuning
of the alloy chemical compositions \cite{chen2018extraordinary,sun2017thermoelectric,wang2020enhanced,guo2020enhanced,saparamadu2020achieving,shuai2016higher, zhang2016designing,
toberer2010electronic,sun2017thermoelectric,sun2017computational}. Many of these materials also show diverse topological properties, ranging from a topological-insulator phase in EuSn$_2$As$_2$ and YbMg$_2$Bi$_2$ \cite{li2019dirac,kundu2022topological}, Dirac semimetal in EuMg$_2$Bi$_2$ \cite{kabir2019observation}, TDS in BaMg$_2$Bi$_2$ \cite{takane2021dirac}, to a Weyl semimetal in  EuCd$_2$As$_2$ \cite{ma2019spin,ma2020emergence}, and topological-crystalline axion insulator in EuIn$_2$As$_2$ \cite{riberolles2021magnetic}. Thus, the 122-Zintl compounds provide versatile platforms for exploring electronic properties and topological phase transitions. Isostructural SmMg$_2$Bi$_2$ is reported to exhibit moderate thermoelectric efficiency \cite{ramirez2015synthesis} and band structure engineering and alloying effects are shown to enhance its efficiency by $\sim$ 50\% \cite{huo2022enhanced,saparamadu2020achieving}. So, for better understanding and tuning the properties of SmMg$_2$Bi$_2$, understanding of the experimental band structure is essential. However, the experimental band structure even for the undoped SmMg$_2$Bi$_2$ is still unknown.

In addition, the valence state of Sm in SmMg$_2$Bi$_2$ is expected to be 2+, where the magnetic susceptibility is expected to follow Van Vleck paramagnetism with $S = 3$, $L = 3$, and $J = 0$. However, the magnetic susceptibility  versus temperature $\chi(T)$ data, reported earlier, show a low-temperature Curie-like behavior \cite{ramirez2015synthesis}, generally observed for systems containing Sm$^{3+}$. The presence of Sm$^{3+}$ in SmMg$_2$Bi$_2$ is yet to be verified experimentally through spectroscopic evidence. Moreover, various mixed-valence Sm-based compounds have been found to exhibit interesting magnetic and transport properties \cite{pakhira2018role,PhysRevLett.123.197203,PhysRevB.91.155151,coey1979magnetic}. It is also expected that this material may exhibit a complex electronic structure if the localized Sm 4$f$ states are hybridized with the conduction electronic states, as found in related materials \cite{kundu2022topological}. Materials with 4$f$ states in the vicinity of $E_\mathrm{F}$ may show heavy-Fermion behavior \cite{PhysRevB.101.045105}. Thus understanding the experimental electronic structure of \smb~is crucial to explain the various interesting physics associated with it.

Here, we report detailed experimental and theoretical studies of the electronic structure and magnetism of SmMg$_2$Bi$_2$ using ARPES, first-principles calculations, and magnetic measurements. Our studies show SmMg$_2$Bi$_2$ to be a mixed-valent Dirac semimetal with a topological electronic structure. The compound belongs to the $Z_2$ topological class having spin-polarized states close to $E_{\rm F}$. Magnetic measurements indicate that the Curie-Weiss-like behavior in $\chi(T)$ originates from the presence of a minor concentration of Sm$^{3+}$ ions, which is also evident in the spectroscopic measurements. We have directly accessed the position of the 4$f$ states and shown that the flat bands and other dispersive states are strongly hybridized.

 \begin{figure*}[ht!]
\centering
\includegraphics[width=17cm]{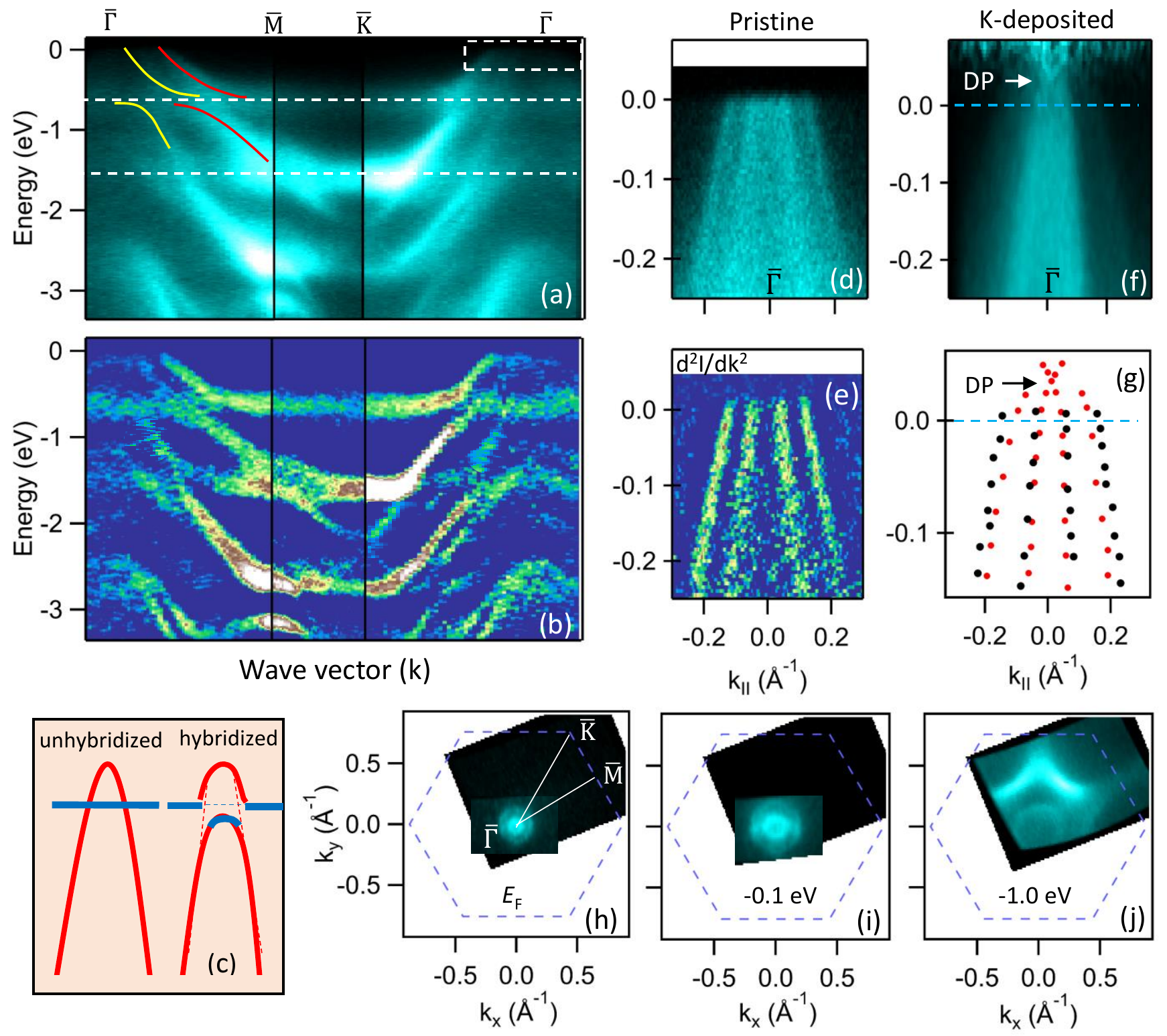}
\caption {Experimental band dispersions and Fermi surface of SmMg$_2$Bi$_2$. (a) ARPES spectrum along the $\bar{\it\Gamma}-\bar{M}-\bar{K}-\bar{\it\Gamma}$ path. A pair of Sm-4$f$ (Sm$^{2+}$) derived flat-bands are highlighted by dashed horizontal lines. Red and yellow curves show the dispersion of two sets of bands that cross the Fermi level and strongly hybridize with the 4$f$ states. (b) Two-dimensional (2D) second derivative of (a) using the procedure as described in Ref.~\cite{zhang2011precise}. (c) A schematic representation of unhybridized (left) and hybridized (right) picture. Red and blue curves represent the dispersing and flat-bands, respectively. (d) High resolution ARPES spectra close to $E_\mathrm {F}$ within the dashed-rectangle shown in (a) at T= 20 K. (e) Second derivative ($d^2I/dk^2$) along the momentum direction of the ARPES spectrum intensity (I) from (d). (f) Same as (d), but after potassium deposition. The data were collected at $\simeq$ 100 K and the intensity was divided by the Fermi-Dirac distribution. The horizontal dashed line is the Fermi energy. The Dirac point (DP) is indicated by an arrow. (g) Band dispersions around $\bar{\it\Gamma}$ for pristine (black) and K-deposited (red) SmMg$_2$Bi$_2$, obtained by tracing the peak position of momentum distribution curves (MDCs). (h)--(j) Fermi surface and constant energy contours of the pristine sample recorded in the second SBZ. The dashed hexagons represent the surface Brillouin zone.}\label{Fig2}
\end{figure*}

\section{Experimental and theoretical details}
\smb\ single crystals were grown using the metal-flux method with a starting composition SmMg$_4$Bi$_6$, similar to an earlier report~\cite{ramirez2015synthesis}. Room-temperature powder x-ray diffraction (XRD) measurements were performed on ground crystals using a Rigaku Geigerflex x-ray diffractometer with Cu-$K_\alpha$ radiation. The XRD data were refined for structural characterization using the Rietveld method with the FULLPROF software package~\cite{Carvajal1993}. The magnetic measurements were performed using a Magnetic-Properties-Measurement System (MPMS) from Quantum Design, Inc., in the $T$ range 1.8 -- 300~K and with magnetic fields ($H$) up to 5.5~T (1~T~$\equiv10^4$~Oe). For the ARPES measurements, the samples were cleaved in-situ under ultra-high vacuum (UHV) just before the measurements. The ARPES experiments were carried out at OASIS-laboratory at Brookhaven National Laboratory (BNL) using a Scienta SES-R4000 electron spectrometer with monochromatized He I$_\alpha$(21.22 eV) radiation (VUV-5k) \cite{Kim2018a}. The total instrumental energy resolution was $\sim$ 10 meV and the angular resolution was better than 0.15$^{\circ}$ and 0.4$^{\circ}$ along and perpendicular to the slit of the analyzer, respectively. Most of the data were taken at {\it T} $\sim$ 20 K. Some of the ARPES data were collected at the 21-ID-1 ESM beamline of the National Synchrotron Light Source II (NSLS-II) with a DA30 Scienta electron spectrometer and a base temperature of 10 K.

First-principles calculations have been carried out using Quantum ESPRESSO \cite{giannozzi2009quantum,giannozzi2017advanced,giannozzi2020quantum}. For the exchange and correlation energy/potential we used the PBEsol functional \cite{perdew2008restoring}. The projected-augmented-wave \cite{PhysRevB.50.17953} method has been used to represent the core electrons. All the calculations have been carried out using the experimental lattice parameters. The cutoff energy for the plane waves and charge density were set to 120 Ry and 520 Ry, respectively. A $k$-mesh of 12$\times$12$\times$10 has been used for the Brillouin Zone integration. The electronic structure calculations have been carried out with  spin-orbit coupling (SOC). Pw2wannier interface and WANNIER90 were used for the construction of the first-principle tight binding Hamiltonian \cite{mostofi2014updated}. Surface-states spectra have been calculated using the WANNIERTOOLS package \cite{wu2018wanniertools}.

\section{Results and discussion}

\subsection{Structural details}
The trigonal crystal structure, room-temperature x-ray diffraction (XRD) pattern, and low-energy electron diffraction (LEED) pattern of SmMg$_2$Bi$_2$ are shown in Figs.~\ref{Fig1}(a), \ref{Fig1}(b), and \ref{Fig1}(c), respectively.  Based on the Zintl concept, the crystal structure can be understood as polyanionic [Mg$_2$Bi$_2$]$^{2-}$ layers aligned along the $c$-axis and separated by trigonal layers of Sm$^{2+}$. The room-temperature powder XRD pattern of ground \smb\ single crystals is shown in Fig.~\ref{Fig1}(b) along with the Rietveld refinement. All Bragg peaks were indexed according to the trigonal ${\rm CaAl_2Si_2}$ crystal structure with space group $P\overline{3}m1$ (No.~164) and hexagonal lattice parameters $a = b = 4.7749(2)$~\AA\ and $c = 7.8406(4)$~\AA. Crystallographic and refinement parameters are shown in Table \ref{TableXRD}. The refined stoichiometry obtained from the Rietveld refinement is ${\rm Sm_{0.98(1)}Mg_{2.05(5)}Bi_{2.01(1)}}$, which is nearly stoichiometric \smb~ to within the error bars.

\begin{table}
\caption{Crystallographic and refinement parameters obtained from the structural analysis of room-temperature powder x-ray diffraction data for \smb .}\label{TableXRD}
 \begin{center}
  \begin{tabular}{ c c c c c c  }
  \hline\hline
  \multicolumn{2}{c}{Lattice parameters} & \multicolumn{4}{c}{}  \\
  \hline
  \multicolumn{2}{c}{{\it a}({\AA})} & \multicolumn{4}{c}{4.7749(2)}  \\
  \multicolumn{2}{c}{{\it c}({\AA})} & \multicolumn{4}{c}{7.8406(4)}  \\
   \multicolumn{2}{c}{V$_{\rm cell}$ ({\AA}$^3$)} & \multicolumn{4}{c}{154.81(1)}  \\
   \multicolumn{2}{c}{Refinement quality} & \multicolumn{4}{c}{}  \\
   \hline
   \multicolumn{2}{c}{$\chi^2$} & \multicolumn{4}{c}{3.02}  \\
   \multicolumn{2}{c}{R$_{\rm Bragg}$ (\%)} & \multicolumn{4}{c}{8.42}  \\
   \multicolumn{2}{c}{R$_{\rm {f}}$ (\%)} & \multicolumn{4}{c}{8.06}  \\
       \multicolumn{2}{c}{Atomic coordinates} & \multicolumn{4}{c}{} \\
       \hline
  Atom & \multicolumn{2}{c}{Wyckoff Symbol} & {\it x} & {\it y} & {\it z}  \\

  Sm & \multicolumn{2}{c}{1\textit{a}} & 0 & 0 & 0 \\
  Mg & \multicolumn{2}{c}{2\textit{d}} & 1/3 & 2/3 & 0.6210(3) \\
  Bi & \multicolumn{2}{c}{2\textit{d}} & 1/3 & 2/3 & 0.2517(2) \\
  \hline\hline
 \end{tabular}
\end{center}

\end{table}

LEED patterns taken from the cleaved SmMg$_2$Bi$_2$ single crystal for two different excitation energies 71 eV and 89 eV are shown in Fig.~\ref{Fig1}(c). The observation of hexagonal patterns confirms that the cleaved surface is the (001) plane. Among six LEED spots, three alternative spots (red circles) are enhanced when probed at 89 eV, reflecting the 3--fold symmetry of the bulk crystal structure \cite{kunduVI3}. Schematics of the hexagonal bulk Brillouin zone and its surface projection (2D dashed hexagon), {\it{i.e.}} surface Brillouin zone (SBZ), are shown in Fig.~\ref{Fig1}(d).

\subsection{Experimental electronic structures}

 Figure~\ref{Fig2} shows the ARPES spectrum along various high-symmetry lines of the SBZ and the Fermi surface of SmMg$_2$Bi$_2$. Figure~\ref{Fig2}(a) shows the ARPES spectrum along the $\bar{\it\Gamma}-\bar{M}-\bar{K}-\bar{\it\Gamma}$ path. Two flat-bands are observed at energies $-$0.65 eV and $-$1.5 eV below $E_\mathrm {F}$. These are the spin-orbit-split Sm$^{2+}$ states. It is also seen that when dispersing bands cross these flat-bands they strongly hybridize and their dispersion changes as highlighted in the schematic representation [Fig.~\ref{Fig2}(c)], where the red and blue curves represent the dispersing and flat-bands, respectively. As a result of hybridization, the orbital character is mixed, the band dispersion changes, and a hybridization gap opens up. Change of band dispersion and hybridization gaps are better visualized in the second derivative plot, Fig.~\ref{Fig2}(b).

 High-resolution ARPES spectra in the vicinity of $E_\mathrm {F}$, around the zone center, is shown in Fig.~\ref{Fig2}(d) and its second derivative in Fig.~\ref{Fig2}(e), respectively. The spectra show two linearly-dispersive hole-like bands crossing the Fermi level. These linearly-dispersing bands form nearly-circular Fermi surfaces, Fig.~\ref{Fig2}(h). The constant-energy contours at $-$0.1 eV and $-$1.0 eV are also shown in Fig.~\ref{Fig2}(i), and \ref{Fig2}(j), respectively. With increasing energy, the inner contour is still nearly circular but the outer contour exhibits hexagonal warping. A very similar low-energy electronic structure is observed in $A$Mg$_2$Bi$_2$ ($A =$ Yb, Ca, Ba) materials \cite{kundu2022topological,takane2021dirac}. By assuming the isotropic spherical shape of the two Fermi surfaces (FS) formed by the two hole-like bands, we estimate the hole concentrations to be $\sim4.7\times$10$^{19}$/cm$^{3}$, which approximately agrees with the value $\sim1\times$10$^{19}$/cm$^{3}$ estimated from transport measurements \cite{saparamadu2020achieving}. The Fermi velocities of the inner and outer hole-like bands are estimated to be 3.4 and 1.9 eV {\AA}, respectively, comparable to those of the topological Dirac semimetals Na$_3$Bi (2.4 eV {\AA}) \cite{liu2014discovery} and BaMg$_2$Bi$_2$ (4.2 eV {\AA}) \cite{takane2021dirac}, implying a potential to achieve a high mobility in SmMg$_2$Bi$_2$.

 To access the unoccupied part of these two sets of hole-like bands, surface potassium deposition was employed while the sample was held at 50 K. After potassium deposition, the ARPES spectrum [Fig.~\ref{Fig2}(f)] collected at 100 K shows a Dirac-like band crossing, just above $E_\mathrm {F}$. We note that in Fig.~\ref{Fig2}(f), the spectrum was divided by the Fermi-Dirac function to better resolve the spectral feature around $E_\mathrm {F}$. The topological origin of this band crossing is discussed in detail in Sec. III.D. Band dispersions extracted from the momentum distribution curves (MDCs) for pristine (black) and K-deposited (red) SmMg$_2$Bi$_2$ are plotted together in Fig.~\ref{Fig2}(g). Upon potassium deposition, the whole band structure appears slightly shifted downward and the Dirac point becomes visible at $\sim$ 30 meV above $E_\mathrm {F}$. These curves superimpose when a downward energy shift of $\sim$ 50 meV is applied to the pristine curve (not shown).

\begin{figure}[ht!]
\centering
\includegraphics[width=8.5cm]{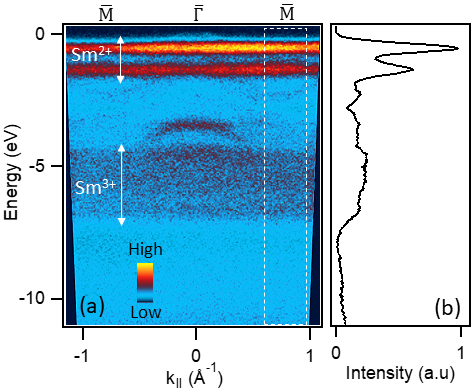}
\caption {Presence of Sm$^{2+}$ and Sm$^{3+}$ in the SmMg$_2$Bi$_2$ spectra. (a) The ARPES intensity recorded at $h\nu =$ 70 eV shows the flat-bands originating from localized Sm 4$f$ states. (b) The energy distribution curve obtained by integrating the photoemission intensity within the dashed rectangle in (a).}\label{Fig3}
\end{figure}

Detailed information about the valence state of Sm can be obtained from photoemission experiments \cite{strisland1997photoemission,pakhira2018role}. In photoemission experiments, when an electron is photo-excited from a partially-filled 4$f^n$ shell, the remaining ($n-$1)4$f$ electrons are left in the various possible multiplet states of the final 4$f^{n-1}$ configuration, which appear at different energies. The ARPES spectrum of SmMg$_2$Bi$_2$ in a larger energy window taken at $h\nu =$ 70 eV to enhance the photoelectron contribution  from Sm-4$f$ states (photo-ionization crosssection of Sm 4$f$ is maximal around $h\nu =$ 70 eV) is shown in Fig.~\ref{Fig3}(a). The flat-bands originating from localized Sm 4$f$ states dominate the spectrum at this photon energy. The two intense flat-bands between $-$0.3 and $-$2 eV originate from Sm$^{2+}$ while the bands at higher energies (between $-$4 and $-$7.5 eV) are due to Sm$^{3+}$ \cite{suga1985resonant,pakhira2018role}. The energy-distribution curve (EDC) is shown in Fig.~\ref{Fig3}(b). The much weaker intensity of Sm$^{3+}$ suggests that Sm$^{2+}$ dominates in SmMg$_2$Bi$_2$. Note that the exact quantification would be difficult as Sm-derived states overlap with other valence states and the ratio of Sm$^{2+}$/Sm$^{3+}$ generally changes significantly with the probing photon energy \cite{Krill,DESY,orlowski2008resonant}. Indeed, the magnetic properties discussed in the following section indicate a concentration of Sm$^{3+}$ of only 2.5(1)\%. The apparent larger concentration of Sm$^{3+}$ in ARPES measurements most likely arises from surface effects such as surface termination, resulting in inhomogeneous defect concentrations between the surface and bulk.

\subsection{Magnetic characterization of SmMg$_2$Bi$_2$}

\begin{figure}[ht]
\includegraphics[width = 3.3in]{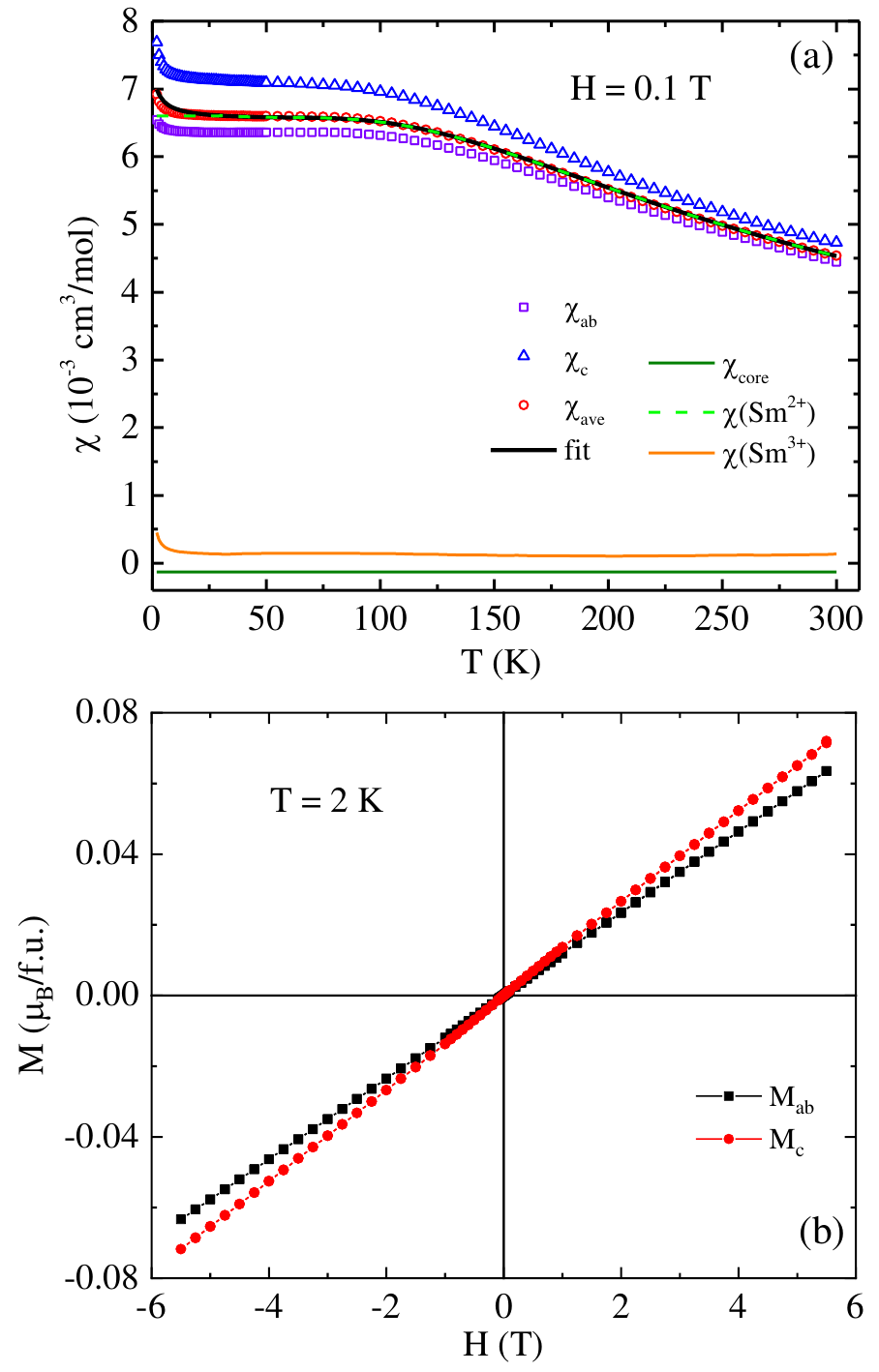}
\caption {(a)~Magnetic susceptibility versus temperature $\chi (T)$ for a \smb\ crystal measured at $H = 0.1$~T with $H \parallel ab$ (open violet squares) and $H \parallel c$ (open blue triangles). The spherically-averaged susceptibility data $\chi_{\rm ave} = (2/3)\chi_{\rm ab} + (1/3)\chi_{\rm c}$ are shown as the open red circles and are fitted by Eqs.~(\ref{Eqs:VanVleck}) as shown by the black curve. The individual contributions of the atomic core diamagnetism, the Curie-Weiss-type susceptibility of the $\sim$ 2.5 mol \% Sm$^{3+}$ spins and the Van Vleck paramagnetic susceptibility of $\sim$ 97.5 mol \% Sm$^{2+}$ spins are also plotted. (b)~Magnetization versus magnetic field $M(H)$ measured at $T = 2$~K for $H \parallel ab$ (black squares) and $H \parallel c$ (red circles).}
\label{Fig4}
\end{figure}

\begin{figure*}[ht!]
\centering
\includegraphics[width=17cm]{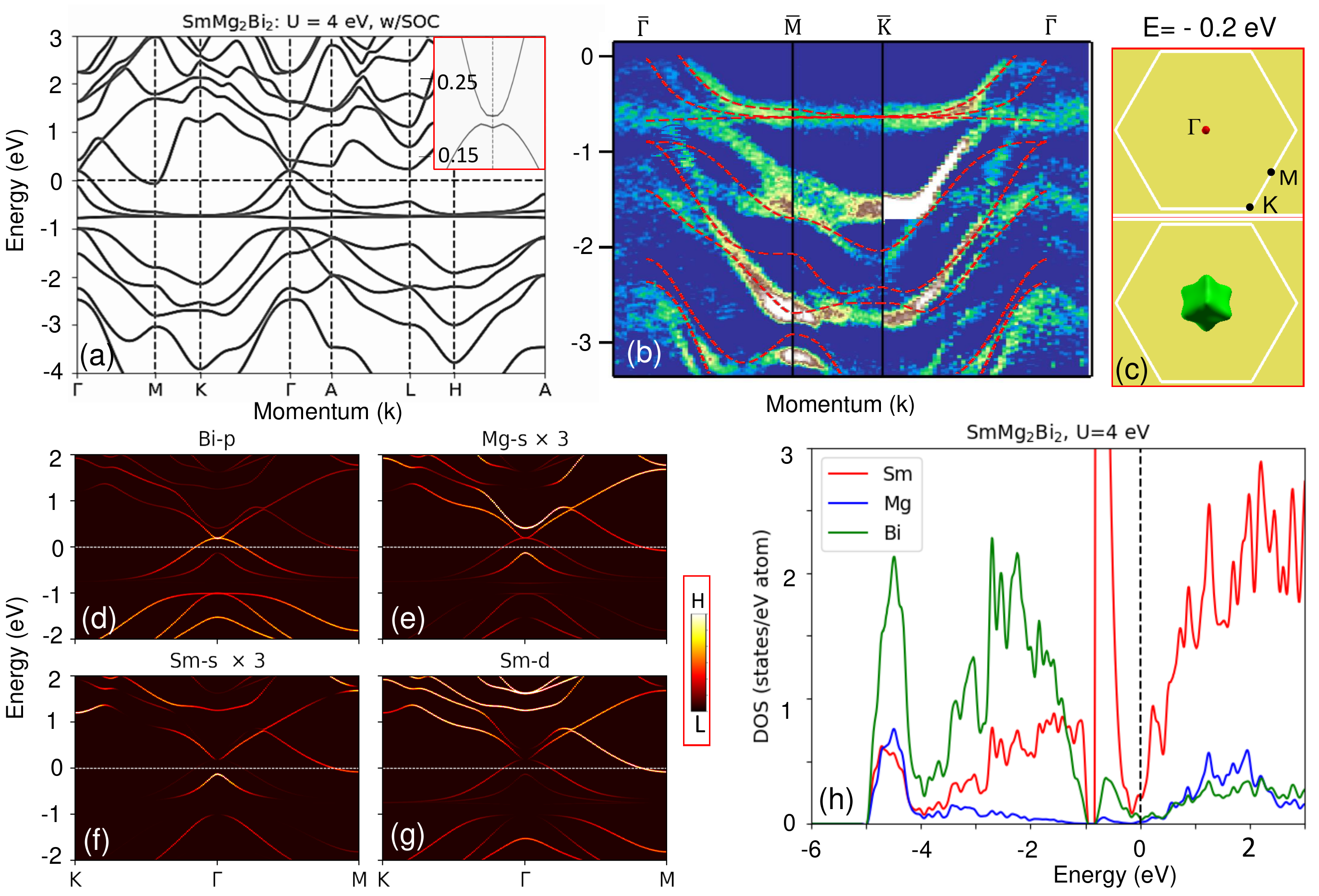}
\caption {Calculated bulk band-structure of SmMg$_2$Bi$_2$. (a) Band dispersions of SmMg$_2$Bi$_2$ with SOC and a Hubbard $U$ = 4 eV. (b) 2D second derivative of ARPES spectrum along the $\bar{\it\Gamma}-\bar{M}-\bar{K}-\bar{\it\Gamma}$ path. Theoretical bulk bands (dashed lines) are superimposed on top. (c) Calculated 3D constant-energy contour at E = -- 0.2 eV viewed perpendicular to the (001) surface. Spherical (top) and hexagonal (bottom) shaped contours are formed by the inner and outer hole-like bands at $\it \Gamma$. (d)-(g) Band dispersions of Bi-$p$, Mg-$s$, Sm-$s$, and Sm-$d$ orbitals with SOC for SmMg$_2$Bi$_2$. The spectral intensity of the Mg-$s$ and Sm-$s$ bands is multiplied by a factor of three to enhance the visibility. (h) Partial density of states (PDOS) plot of SmMg$_2$Bi$_2$, sum over the both spin channels. }\label{Fig5}
\end{figure*}
The magnetic susceptibility $\chi(T)$ data of a \smb\ single crystal measured in $H = 0.1$~T for both in-plane ($H \parallel ab$) and out-of-plane ($H \parallel c$) field directions are shown in Fig.~\ref{Fig4}(a). As reported earlier~\cite{ramirez2015synthesis}, the $\chi(T)$ exhibits a plateau in the $T$~range 20--100~K followed by a Curie-like upturn in low~$T$\@.

As per Zintl's theory for stoichiometric SmMg$_2$Bi$_2$, the Sm valency should be 2+, taking $3-$ for Bi and 2+ for Mg. The electron configuration of Sm$^{2+}$ is $4f^6$, leading by Hund's rules to the $^7F_0$ state with quantum numbers
\bse
\bea
S =3,~L=3,~J=0 \quad({\rm for~Sm}^{2+}).\label{Eq:Sm+2Pars}
\eea
Thus this ion has an orbital (Van Vleck) contribution but no spin contribution to its $\chi(T)$.  On the other hand, the Sm$^{+3}$ ion has the electron configuration $4f^5$ with state $^6H_{5/2}$ and quantum numbers
\bea
S=5/2,~L=5,~J=5/2\quad({\rm for~Sm}^{3+}),\label{Eq:Sm+3Pars}
\eea
\ese
and hence has a spin susceptibility. Our photoemission measurements discussed earlier indicate the presence of mainly Sm$^{2+}$ with a minor amount of Sm$^{3+}$ cations in the system, where the latter are presumed associated with magnetic defects.

Therefore, the molar $\chi(T)$ data for \smb\ are analyzed considering both Sm$^{2+}$ and Sm$^{3+}$ contributions as follows
\bea
\chi(T) = \chi_{\rm core} + n_{\rm frac} \chi({\rm Sm^{3+}})(T) \nonumber\\
 + (1-n_{\rm frac}) \chi({\rm Sm^{2+}})(T),
\label{Eq:chiBoth}
\eea

\noindent where the isotropic atomic core diamagnetism $\chi_{\rm core}$ is estimated to be $\chi_{\rm core}\sim -2\times 10^{-4}\,{\rm cm^3/mol}$~\cite{Selwood1956} and $n_{\rm frac}$ is the molar fraction of ${\rm Sm^{3+}}$ spins.  The resulting molar Curie-Weiss law spin susceptibility of the Sm$^{3+}$ spins  is
\bse
\label{Eqs:Sm3+chi}
\bea
\chi({\rm Sm^{3+}})(T) &=& \frac{C}{T-\theta_{\rm p}},\\
C &=& \frac{N_Ag^2\mu_{\rm B}^2J(J+1)}{3k_{\rm B}},\label{Eq:C1}
\eea
\ese
where $C$ is the Sm$^{3+}$ molar Curie constant, $\theta_{\rm p}$ is the paramagnetic Weiss temperature, $N_{\rm A}$ is Avogadro's number, $g = 2/7$ is the Land\'e $g$~factor of Sm$^{3+}$, $\mu_{\rm B}$ is the Bohr magneton, $k_{\rm B}$ is Boltzmann's constant, and $J=5/2$ from Eq.~(\ref{Eq:Sm+3Pars}).  Then Eq.~(\ref{Eq:C1}) gives the value of the Sm$^{3+}$ molar Curie constant as
\bea
C = 0.08930\,\frac{{\rm cm^3\,K}}{{\rm mol\,Sm}^{3+}},
\label{Eq:C}
\eea
where we have ignored the Van Vleck paramagnetism of Sm$^{3+}$ ($L$ = 5).
The spin of Sm$^{2+}$ is $S=3$ from Eq.~(\ref{Eq:Sm+2Pars}), yielding a multiplicity $2S+1 =  7$ for the $J$ states \mbox{$J = 0$, 1, $\ldots$, 6.}  According to the Van Vleck theory, the orbital molar magnetic susceptibility of Sm$^{2+}$ is given by~\cite{VanVleck1932, Takikawa2010}
\bse
\label{Eqs:VanVleck}
\bea
\chi ({\rm Sm^{2+}}) &=&  \frac{N_{\rm A}\sum_{J = 0}^{J = 6}\bigg[\frac{g^2\mu_{\rm B}^2J(J+1)}{3k_{\rm B}T}+\alpha_J\bigg](2J+1)e^{-\frac{E_J}{k_{\rm B}T}}}{\sum_{J = 0}^{J = 6}(2J+1)e^{-E_J/k_{\rm B}T}}\nonumber\\
&=& \frac{\sum_{J = 0}^{J = 6}\bigg[\frac{C}{T}+N_{\rm A}\alpha_J\bigg](2J+1)e^{-\frac{E_J}{k_{\rm B}T}}}{\sum_{J = 0}^{J = 6}(2J+1)e^{-E_J/k_{\rm B}T}},
\label{Eq:3}
\eea
where
\bea
E_J &=& \frac{\lambda}{2}[J(J+1)-L(L+1)-S(S+1)],\\
E_J&-&E_{J-1} = \lambda J,
\label{Eq:6}
\eea
$\lambda$ is the spin-orbit coupling energy and the values of $S,~L,$ and~$J$ for Sm$^{+3}$ are given in Eq.~(\ref{Eq:Sm+3Pars}).  The quantity $\alpha_J$ is given by
\bea
\alpha_J &=& \frac{\mu_{\rm B}^2}{6(2J+1)}\bigg[\frac{F_{J+1}}{E_{J+1}-E_J}-\frac{F_J}{E_J-E_{J-1}}\bigg]
\label{Eq:4}
\eea
with
\bea
F_J = \frac{[(L+S+1)^2-J^2][J^2-(S-L)^2]}{J}.
\label{Eq:5}
\eea
\ese

The fit of the $\chi_{\rm ave}(T)$ data in Fig.~\ref{Fig4}(a) by Eqs.~(\ref{Eqs:VanVleck}) is shown as the black curve in Fig.~\ref{Fig4}(a).  The fitted parameters are $n_{\rm frac} = 0.025(1)$, $\theta_{\rm p} = -2.9(2)$~K, and $\lambda_{\rm ave}/k_{\rm B} = 438.6(5)$~K\@.   The value of $\alpha_{\rm ave}$ indicates the presence of $2.5$\% mole fraction of magnetic Sm$^{3+}$ ions in the crystal. The finite negative value of $\theta_{\rm p}$ indicates possible antiferromagnetic correlations among the dilute Sm$^{3+}$ spins. The value of $\lambda_{\rm ave}/k_{\rm B}$ is nearly same as the average value
\bea
\lambda_{\rm ave}/k_{\rm B} = [\lambda(H\parallel c)+2\lambda(H\parallel ab)]/(3k_{\rm B}) = 436~{\rm K}
\eea
deduced in Ref.~\cite{ramirez2015synthesis}. As the XRD data do not reveal the presence of any additional phases in the crystal, the Sm$^{3+}$ ions may arise from nonstoichiometry of \smb, as suggested from the XRD refinement.

The isothermal magnetization versus field $M(H)$ data measured at $T = 2$~K for $H \parallel ab$ and $H \parallel c$ with $0\leq H\leq 5.5$~T are shown in Fig.~\ref{Fig4}(b). The $M(H)$ behavior is anisotropic, consistent with the data in Fig.~\ref{Fig4}(a).  The data for each direction are linear at higher fields, while a very weak nonlinearity is observed at lower fields, plausibly due to the presence of dilute Sm$^{3+}$ ions and/or magnetic defects.  No magnetic hysteresis is observed.

\subsection{Comparison of band dispersions: Theory vs. ARPES}

Figure~\ref{Fig5} shows the calculated bulk electronic structure along various high-symmetry paths for SmMg$_2$Bi$_2$. Figure~\ref{Fig5}(a) represents the band dispersions of SmMg$_2$Bi$_2$ including spin-orbit coupling (SOC) and a Hubbard $U$ parameter of 4 eV. The Hubbard $U$ is used to describe the localized Sm-4$f$ states and to match the energy position of the observed flat-bands in the ARPES spectra. The valence and conduction bands apparently touch each other at the $\it \Gamma$ point and form a Dirac-like band crossing, consistent with our ARPES results as discussed earlier in Sec. III.A. A closer look around the $\it \Gamma$ point [inset of Fig.~\ref{Fig5}(a)] reveals that a small energy gap of about $\sim$ 2 meV is present between the valence and conduction bands and the shape of these bands suggest that there might be a band inversion. Besides, a strong hybridization between the $4f$ states and highly-dispersive bands is also observed where they cross each other at around -- 0.8 eV. The calculated constant energy contour at $-$ 0.2 eV in the 3D BZ is shown in Fig.~\ref{Fig5}(c). The inner and outer hole-like bands at the $\Gamma$ point form spherical (top) and wrapped hexagonal-shaped (bottom) contours, similar to those observed in our ARPES experiments. A direct comparison between ARPES spectrum and the theoretical band dispersions is shown in Fig.~\ref{Fig5}(b). Although most of the experimental features are captured well by the calculations, the calculated spectrum shows only one flat-band, while two flat-bands are observed experimentally within the 0 to $-$ 4.0 eV energy window. By systematically changing the $U$ values in the calculations (not shown), we found that with increasing $U$, the Sm 4$f_{7/2}$ states became unoccupied and shifted far above $E_\mathrm {F}$ [outside the energy window of Fig.~\ref{Fig5}(a)] and the states below $E_\mathrm {F}$ are the Sm 4$f_{5/2}$ states. This is in sharp contrast to our experimental observation where both the Sm 4$f_{5/2}$ and Sm 4$f_{7/2}$ states are completely occupied. This problem is intrinsic to the DFT + $U$ calculations and originates from the improper treatment of partially-filled 4$f$ states in the DFT + $U$ framework \cite{noorafshan2018lda,campo2010extended}.

We note that to match the experimental bands, a rigid upward energy shift of $\sim$ 0.2 eV was applied to the theoretical bands in Fig.~\ref{Fig5}(b), suggesting that the material is hole-doped. Our XRD analysis also supports this scenario and indicates that hole-doping likely originates from intrinsic vacancies on the $A$ sites. A similar trend is also found in materials such as $A$Mg$_2$Bi$_2$ ($A=$ Yb, Ca, Eu) \cite{may2011structure,shuai2016higher}.

To understand the atomic and orbital character of the bands, orbital-resolved band dispersions from the Bi-$p$, Mg-$s$, Sm-$s$, and Sm-$d$ states are shown in Figs.~\ref{Fig5}(d)--\ref{Fig5}(g). It can be seen that the outer hole-like valence band has dominant Bi-$p$ character while the inner one has dominant Mg-$s$ and Sm-$s$ character. The electron-like band at the $M$-point around $E_\mathrm {F}$ is dominated by Sm-$d$ character. However, it is also evident that all the bands have mixed orbital character as they strongly hybridize with each other. We note that small contributions of other orbitals such as Bi-$s$, Mg-$p$, and Sm-$p$ are also present around $E_\mathrm {F}$. Figure~\ref{Fig5}(h) shows the partial densities of states (PDOS) of SmMg$_2$Bi$_2$. Near $E_\mathrm {F}$, the DOS is dominated by Sm and Bi atoms. A strong peak from the Sm atoms at $\sim-$ 0.65 eV originates from the Sm-4$f$ states.
\begin{figure}[ht!]
\centering
\includegraphics[width=8.5cm]{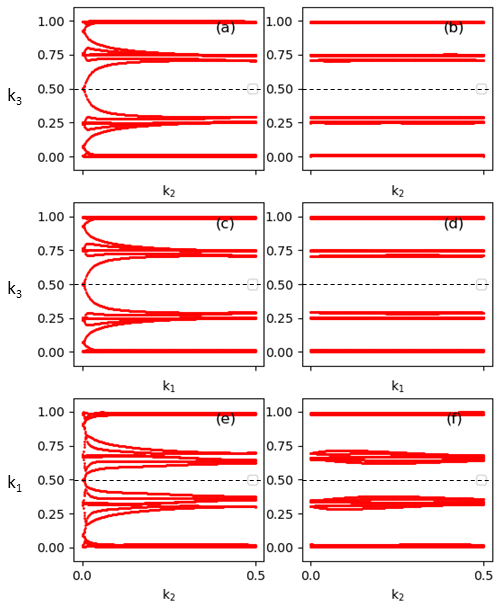}
\caption {Wilson-loop evolution at six TRS-invariant momentum planes and $Z_2$ numbers. (a) $k_1$ = 0.0, $k_2-k_3$ plane, $Z_2$ = 1. (b) $k_1$ = 0.5, $k_2-k_3$ plane, $Z_2$ = 0. (c) $k_2$ = 0.0, $k_1-k_3$ plane, $Z_2$ = 1. (d) $k_2$ = 0.5, $k_1-k_3$ plane, $Z_2$ = 0. (e) $k_3$ = 0.0, $k_2-k_1$ plane, $Z_2$ = 1. (f) $k_3$ = 0.5, $k_2-k_1$ plane, $Z_2$ = 0. Wilson loop crosses the reference line (dashed horizontal lines) an odd number of times for (a), (c) and (e) and an even number of times for (b), (d) and (f), resulting in $Z_2$ = 1 and $Z_2$ = 0, respectively.}\label{Fig6}
\end{figure}

To verify the nontrivial topology of the low-energy electronic states as hinted by the shape of the valence and conduction bands near the zone center [inset of Fig.~\ref{Fig5}(a)], we have calculated the $Z_2$ topological numbers using the Wilson loop (Wannier charge center) method \cite{yu2011equivalent} for the six time-reversal-invariant momentum planes, shown in Figs.~\ref{Fig6}(a)--\ref{Fig6}(f). Six TRS planes are (a) $k_1$ = 0.0, (b) $k_1$ = 0.5, (c) $k_2$ = 0.0, (d) $k_2$ = 0.5, (e) $k_3$ = 0.0, and (f) $k_3$ = 0.5, where $k_1$ , k$_2$, and $k_3$ are in units of reciprocal lattice vector. It is well established that if the Wilson loop crosses any reference horizontal line an odd (even) number of times, then $Z_2=$ 1 ($Z_2=$ 0). Considering this concept, our results show $Z_2=$ 1 for $k_1$ = 0.0, $k_2$ = 0.0, and $k_3$ = 0.0 planes and $Z_2=$ 0 for $k_1$ = 0.5, $k_2$ = 0.5, and $k_3$ = 0.5 planes. So, the $Z_2$ topological numbers of this system $v_0$;$(v_1v_2v_3)$ $=$ 1;(000) indicate that SmMg$_2$Bi$_2$ is a topological material, where $v_0 =$ [$Z_2$($k_i = $ 0)$+$$Z_2$($k_i = $ 0.5)] mod 2 and $v_i=Z_2$($k_i = $ 0.5).

\begin{figure}[ht!]
\centering
\includegraphics[width=8.5cm]{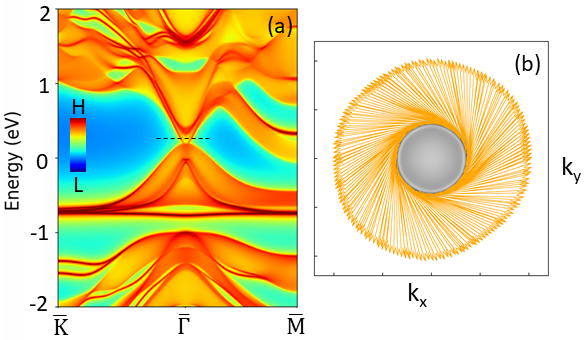}
\caption {Calculated surface-state spectrum and spin-texture of SmMg$_2$Bi$_2$. (a) Surface-state spectrum from a (001) semi-infinite slab of a Mg-terminated surface. The spectral brightness indicates the charge density integrated over the top six atomic layers, $i.e.$, a high intensity represents a higher surface-state contribution. (b) Helical spin-texture at an energy cut as shown by the dashed line in (a). }\label{Fig7}
\end{figure}

Since the bulk band-structure calculations show evidence of band inversion and the calculated $Z_2$ numbers confirm that the system is topological, we have calculated the surface-state spectrum to look for any evidence of topological surface states. The surface electronic structure calculated from a (001) semi-infinite slab is shown in Fig.~\ref{Fig7}(a). Various surface states (high intensity states) can be seen but most of them appear at the edges of the projected bulk bands. A Dirac-like band crossing is also observed at $\bar{\it \Gamma}$, just above $E_\mathrm {F}$. The Dirac-like bands show helical spin-texture [Fig.~\ref{Fig7}(b)], similar to those observed in topological insulators. Thus, the theoretical results suggest that SmMg$_2$Bi$_2$ is an extremely narrow-gap topological insulator. Further, the ARPES results show a gapless Dirac-like spectrum, similar to that observed in the Dirac semimetal BaMg$_2$Bi$_2$ \cite{takane2021dirac}. However, the detection of such a small gap (if present in the real system!) is outside the limit of our experimental resolution. These results together suggest that SmMg$_2$Bi$_2$ is in proximity to a quantum-critical point of a topological Dirac semimetallic phase.

\subsection{Summary}

In summary, we have investigated the electronic and magnetic properties of SmMg$_2$Bi$_2$ using a combination of ARPES, DFT, and magnetic measurements. The theoretical and ARPES results suggest that SmMg$_2$Bi$_2$ is in close proximity to a topological Dirac semimetallic phase. The ARPES results show a Dirac-like band crossing at the zone-center near the Fermi level ($E_\mathrm {F}$), which is further confirmed by a first-principle calculation. Theoretical results also reveal that SmMg$_2$Bi$_2$ hosts spin-polarized states around $E_\mathrm {F}$ which could show some interesting spin-dependent properties. It is also shown that the valence state of Sm in this compound is dominated by a divalent (Sm$^{2+}$) contribution with a small admixture of trivalent Sm (Sm$^{3+}$). The magnetic measurements suggest that the low-temperature upturn in the susceptibility arises due to the presence of paramagnetic Sm$^{3+}$ in this system. In recent years, SmMg$_2$Bi$_2$-based systems have gained increased attention due to their improved thermoelectric performance, where band-structure engineering has been proposed to play a critical role. Thus our detailed spectroscopic study should lead to a better understanding and tuning of the transport and thermoelectric properties of SmMg$_2$Bi$_2$-based systems. Furthermore, by {\it in-situ} doping, we have been able to tune to the chemical potential of this compound, yielding a flexible platform for exploring exotic physical phenomena that should be observable when chemical/gate doping is varied.

\section*{ACKNOWLEDGMENTS}

This work was supported by the US Department of Energy, office of Basic Energy Sciences, Contract No. DE-SC0012704. The research at Ames National Laboratory was supported by the U.S. Department of Energy, Office of Basic Energy Sciences, Division of Materials Sciences and Engineering. Ames National Laboratory is operated for the U.S. Department of Energy by Iowa State University under Contract No. DE-AC02-07CH11358. This work was also supported in part by the Center for Spintronics Research Network, Tohoku University.

\end{document}